\documentstyle[amstex,amssymb,12pt]{article}

\catcode`\@=11
\def\numberbysection{\@addtoreset{equation}{section}
        \def\theequation{\thesection.\arabic{equation}}}
\numberbysection

\begin{document}

\newlength{\lno} \lno0.5cm \newlength{\len} \len=\textwidth%
\addtolength{\len}{-\lno}

\setcounter{page}{0}

\baselineskip7mm \newpage \setcounter{page}{0}

\begin{titlepage}     
\vspace{1.5cm}
\begin{center}
{\Large\bf  $D_{n+1}^{(2)} $ Reflection K-Matrices  }\\
\vspace{1cm}
{\large A. Lima-Santos }\footnote{e-mail: dals@df.ufscar.br} \\
\vspace{1cm}
{\large \em Universidade Federal de S\~ao Carlos, Departamento de F\'{\i}sica \\
Caixa Postal 676, CEP 13569-905~~S\~ao Carlos, Brasil}\\
\end{center}
\vspace{2.5cm}

\begin{abstract}
We investigate the possible regular solutions of the boundary Yang-Baxter
equation for the  vertex models associated to the $D_{n+1}^{(2)}$ affine Lie algebra.
We have classified them in terms of three types of $K$-matrices. The first one have
$n+2$ free parameters and all the matrix elements are non-null. The second  solution is
given by a block diagonal matrix with  just one free parameter. It turns out that for $n$ even
there exists a third class of  $ K$-matrix without free parameter.
\end{abstract}

\vfill
\begin{center}
\small{\today}
\end{center}
\end{titlepage}

\baselineskip6mm

\newpage{}

\section{{}Introduction}

Recently there has been a lot of efforts in introducing boundaries into
integrable systems for possible applications to condensed matter physics and
statistical system with non-periodic boundary conditions. The bulk Boltzmann
weights of an exactly solvable lattice system are usually the non-null
matrix elements of a $R$-matrix $R(u)$ which satisfies the Yang-Baxter
equation. The boundaries entail new physical quantities called reflection
matrices which depend on the boundary properties.

By considering systems on a finite interval with independent boundary
conditions at each end, we have to introduce reflection matrices to describe
such boundary conditions. Integrable models with boundaries can be
constructed out of a pair of reflection $K$-matrices $K_{\pm }(u)$ in
addition to the $R$-matrix. $K_{+}(u)$ and $K_{-}(u)$ describe the effects
of the presence of boundaries at the left and the right ends, respectively.

Integrability of open chains in the framework of the quantum inverse
scattering method was pioneered by Sklyanin. In reference \cite{Sklyanin},
Sklyanin has used his formalism to solve, via algebraic Bethe ansatz, the
open spin-$1/2$ chain with diagonal boundary terms. This \ model had already
been solved via coordinate Bethe ansatz by Alcaraz {\it et al} \cite{Alcaraz}%
.

The Sklyanin original formalism was extended to more general systems by
Mezincescu and Nepomechie in \cite{Mezincescu1}, where is assumed that for a
regular $R$-matrix satisfying the following properties 
\begin{eqnarray}
{\em PT-symmetry} &:&P_{12}R_{12}(u)P_{12}=R_{21}(u),  \nonumber \\
{\em unitarity} &:&R_{12}(u)R_{21}(-u)=1,  \nonumber \\
{\em crossing\ unitarity} &:&R_{12}(u)=(U\otimes 1)R_{12}^{{\rm t}%
_{2}}(-u-\rho )(U\otimes 1)^{-1},  \label{int.1}
\end{eqnarray}%
one can derive an integrable open chain Hamiltonian 
\begin{equation}
H=\sum_{k=1}^{N-1}H_{k,k+1}+\frac{1}{2}(\left. \frac{dK_{-}(u)}{du}\right|
_{u=0}\otimes 1)+\frac{{\rm tr}_{0}\overset{0}{K}_{+}(0)H_{N,0}}{{\rm tr}%
K_{+}(0)},  \label{int.2}
\end{equation}%
where $R_{12}=R\otimes 1$, $R_{23}=1\otimes R$, etc. $P=R(0)$ is the
permutation matrix and the two-site bulk Hamiltonian $H_{k,k+1}$ is given by 
$H_{k,k+1}=P_{k,k+1}\left[ dR_{k,k+1}(u)/du\right] _{u=0}$.

The matrix $K_{-}(u)$ satisfies the right boundary Yang-Baxter equation,
also known as the reflection equation ({\small RE})%
\begin{equation}
R_{12}(u-v)\overset{1}{K_{-}}(u)R_{21}(u+v)\overset{2}{K_{-}}(v)=\overset{2}{%
K_{-}}(v)R_{12}(u+v)\overset{1}{K_{-}}(u)R_{21}(u-v)  \label{int.3}
\end{equation}%
which governs the integrability at boundary for a given bulk theory. \
Similar equation should also hold for the matrix $K_{+}(u)$ at the opposite
boundary. However, for the case of models whose matrix $R(u)$\ satisfies (%
\ref{int.1}), one can show that the corresponding quantity 
\begin{equation}
K_{+}(u)=K_{-}^{{\rm t}}(-u-\rho )M,\qquad M=U^{{\rm t}}U=M^{{\rm t}},
\label{int.4}
\end{equation}%
satisfy the left {\small RE}. \ Here $\rho $ is a crossing parameter and $U$
is a crossing matrix both being specific to each model \cite{Bazhanov,
Schadrikov}. ${\rm t}_{i}$ stands for the transposition taken in the $i$-%
{\it th} space and {\rm tr}$_{0}$ is the trace taken in the auxiliary space.

Due to the significance of the {\small RE} in the construction of integrable
models with open boundaries, a lot of work has been directed to the study  %
\cite{Cherednik,deVega,Mezincescu2,Inami} and classification \cite%
{Batchelor, Lima, Liu} of  $K$-matrices. 

In spite of all these works, there is an interesting vertex model based on
the non-exceptional $D_{n+1}^{(2)}$ Lie algebra for which, until recently,
little was known about the solution of the corresponding {\small RE} \cite%
{MG}.

While the investigation of particular solutions has been made to a number of
lattice models, Batchelor {\it at al} \cite{Batchelor} have derived diagonal
solutions of the {\small RE} for face and vertex models associated with
several affine Lie algebras, \ the classification of all possible $K$%
-matrices has been a harder problem. However, recently we have proposed a
method which allowed to classify all possible $K$-matrices solutions for $19$%
-vertex models\cite{Lima}. 

In this paper we would to demonstrate that this technique is indeed much
more general and can be used, in principle, to study all vertex models based
on the Lie algebras \cite{Bazhanov, Schadrikov}. In order to show that we
choice the $D_{n+1}^{(2)}$ vertex model since it is seen as the more
difficult case among the non-exceptional cases, as argued recently by
Martins and Guan \cite{MG}.

We have organized this paper as follows. In section $2$ we present the $%
D_{n+1}^{(2)}$reflection equations and in section $3$ their solutions are
derived and classified in three types. The last section is reserved for the
conclusion. The $D_{2}^{(2)}$ type-I \ solution is presented in appendix.

\section{The D$_{n+1}^{(2)}$ reflection equations}

The $R$-matrix for the vertex models associated to the $D_{n+1}^{(2)}$
affine Lie algebra as presented by Jimbo in \cite{Jimbo} has the form 
\begin{eqnarray}
R &=&\sum_{i,j\neq n+1,n+2}a_{ij}\ E_{ij}\otimes E_{i^{\prime }j^{\prime
}}+a_{1}\sum_{i\neq n+1,n+2}E_{ii}\otimes E_{ii}+a_{2}\sum\begin{Sb} i\neq
j,j^{\prime }  \\ i\ \text{{\rm or} }j\neq n+1,n+2  \end{Sb}  E_{ii}\otimes
E_{jj}  \nonumber \\
&&+a_{3}\sum\begin{Sb} i<j,i\neq j^{\prime }  \\ i,j\neq n+1,n+2  \end{Sb} 
E_{ij}\otimes E_{ji}+a_{4}\sum\begin{Sb} i>j,i\neq j^{\prime }  \\ i,j\neq
n+1,n+2  \end{Sb}  E_{ij}\otimes E_{ji}  \nonumber \\
&&+a_{5}\sum\begin{Sb} i<n+1  \\ j\neq n+1,n+2  \end{Sb}  \left(
E_{ij}\otimes E_{ji}+E_{j^{\prime }i^{\prime }}\otimes E_{i^{\prime
}j^{\prime }}\right) +a_{6}\sum\begin{Sb} i>n+2  \\ j\neq n+1,n+2  \end{Sb} 
\left( E_{ij}\otimes E_{ji}+E_{j^{\prime }i^{\prime }}\otimes E_{i^{\prime
}j^{\prime }}\right)  \nonumber \\
&&+a_{7}\sum\begin{Sb} i<n+1  \\ j\neq n+1,n+2  \end{Sb}  \left(
E_{ij}\otimes E_{j^{\prime }i}+E_{j^{\prime }i^{\prime }}\otimes
E_{i^{\prime }j}\right) +a_{8}\sum\begin{Sb} i>n+2  \\ j\neq n+1,n+2 
\end{Sb}  \left( E_{ij}\otimes E_{j^{\prime }i}+E_{j^{\prime }i^{\prime
}}\otimes E_{i^{\prime }j}\right)  \nonumber \\
&&+\sum\begin{Sb} i\neq n+1,n+2  \\ j=n+1,n+2  \end{Sb}  \left( b_{i}^{+}%
\left[ E_{ij}\otimes E_{i^{\prime }j^{\prime }}+E_{j^{\prime }i^{\prime
}}\otimes E_{ji}\right] +b_{i}^{-}\left[ E_{ij}\otimes E_{i^{\prime
}j}+E_{ji^{\prime }}\otimes E_{ji}\right] \right)  \nonumber \\
&&+\sum_{i=n+1,n+2}\left[ c^{+}E_{ii}\otimes E_{i^{\prime }i^{\prime
}}+c^{-}E_{ii}\otimes E_{ii}+d^{+}E_{ii^{\prime }}\otimes E_{i^{\prime
}i}+d^{-}E_{ii^{\prime }}\otimes E_{ii^{\prime }}\right]  \label{re.1}
\end{eqnarray}%
where $E_{ii}$ denotes the elementary $2n+2$ by $2n+2$ matrices ($\delta
_{ia}\delta _{ib}$) and the Boltzmann weights with functional dependence on
the spectral parameter $u$ are given by 
\begin{eqnarray}
a_{1} &=&({\rm e}^{2u}-q^{2})({\rm e}^{2u}-q^{2n}),\qquad \ a_{2}=q({\rm e}%
^{2u}-1)({\rm e}^{2u}-q^{2n}),  \nonumber \\
a_{3} &=&-(q^{2}-1)({\rm e}^{2u}-q^{2n}),\qquad \ a_{4}={\rm e}^{2u}a_{3}, 
\nonumber \\
a_{5} &=&\frac{1}{2}({\rm e}^{u}+1)a_{3},\ \qquad \qquad \quad \ \ \ \!a_{6}=%
\frac{1}{2}({\rm e}^{u}+1){\rm e}^{u}a_{3},  \nonumber \\
a_{7} &=&-\frac{1}{2}({\rm e}^{u}-1)a_{3},\qquad \qquad \quad \ a_{8}=\frac{1%
}{2}({\rm e}^{u}-1){\rm e}^{u}a_{3},  \label{re.2}
\end{eqnarray}%
and for $i,j\neq n+1,n+2$%
\begin{equation}
a_{ij}=\left\{ 
\begin{array}{c}
(q^{2}{\rm e}^{2u}-q^{2n})({\rm e}^{2u}-1)\qquad \ \qquad \qquad \ \ \ \ \ \
\ \ \ \ \ \ \qquad \qquad (i=j) \\ 
(q^{2}-1)\ \left( q^{2n+\overset{\_}{i}-\overset{\_}{j}}({\rm e}%
^{2u}-1)-\delta _{ij^{\prime }}({\rm e}^{2u}-q^{2n})\right) \qquad \ \qquad
(i<j) \\ 
(q^{2}-1){\rm e}^{2u}\left( q^{\overset{\_}{i}-\overset{\_}{j}}({\rm e}%
^{2u}-1)-\delta _{ij^{\prime }}({\rm e}^{2u}-q^{2n})\right) \qquad \ \
\qquad (i>j)%
\end{array}%
\right.  \label{re.3}
\end{equation}%
\begin{equation}
b_{i}^{\pm }=\left\{ 
\begin{array}{c}
\pm \ q^{i-1/2}\ (q^{2}-1)\ ({\rm e}^{2u}-1)\ ({\rm e}^{u}\pm q^{n})\qquad
\qquad \qquad \ \ \ \ (i<n+1) \\ 
q^{i-n-5/2}\ (q^{2}-1)\ ({\rm e}^{2u}-1)\ {\rm e}^{u}({\rm e}^{u}\pm
q^{n})\qquad \ \qquad \qquad (i>n+2)%
\end{array}%
\right.  \label{re.4}
\end{equation}%
\begin{equation}
c^{\pm }=\pm \frac{1}{2}(q^{2}-1)\ (q^{n}+1)\ {\rm e}^{u}({\rm e}^{u}\mp 1)\
({\rm e}^{u}\pm q^{n})+q\ ({\rm e}^{2u}-1)\ ({\rm e}^{2u}-q^{2n}),
\label{re.5}
\end{equation}%
\begin{equation}
d^{\pm }=\pm \frac{1}{2}(q^{2}-1)\ (q^{n}-1)\ {\rm e}^{u}({\rm e}^{u}\pm 1)\
({\rm e}^{u}\pm q^{n}),  \label{re.6}
\end{equation}%
with special attention to the notation%
\begin{equation}
\overset{\_}{i}=\left\{ 
\begin{array}{c}
i+1\ ,\qquad \quad \quad (i<n+1) \\ 
n+\frac{3}{2},\quad \ (i=n+1,n+2) \\ 
i-1\ ,\qquad \quad \quad (i>n+2)%
\end{array}%
\right. \qquad {\rm and}\qquad \ i^{\prime }=2n+3-i.  \label{re.7}
\end{equation}%
Here $q={\rm e}^{-2\eta }$ denotes an arbitrary parameter.

For each value of $n$ the matrix $R(u)$ is regular $R(0)=f(0)P$ and satisfy 
{\small PT}-symmetry and unitarity 
\begin{equation}
R_{21}(u)=P_{12}R_{12}(u)P_{12},\qquad R_{12}(u)R_{21}(-u)=f(u)f(-u){\bf 1,}
\label{re.8}
\end{equation}%
where%
\begin{equation}
f(u)=({\rm e}^{2u}-q^{2})({\rm e}^{2u}-q^{2n}).  \label{re.9}
\end{equation}%
After we normalize the Boltzmann weights by a factor $q^{n+1}{\rm e}^{2u}$,
the crossing-unitarity symmetry 
\begin{equation}
R_{12}(u)=(U\otimes 1)R_{12}^{{\rm t}_{2}}(-u-\rho )(U\otimes 1)^{-1},
\label{re.10}
\end{equation}%
holds with the crossing matrices $U$ and crossing parameters $\rho $,
respectively given by 
\begin{equation}
U_{ij}=\delta _{i^{\prime }j}\ q^{(\overset{\_}{i}-\overset{\_}{j}%
)/2},\qquad \rho =-n\ln q=2n\eta  \label{re.11}
\end{equation}

Regular solutions of the {\small RE} mean that the matrix $K_{-}(u)$ in the
form 
\begin{equation}
K_{-}(u)=\sum_{i,j=1}^{2n+2}k_{ij}(u)\ E_{ij}  \label{re.12}
\end{equation}%
satisfies the condition 
\begin{equation}
k_{ij}(0)=\delta _{ij},\quad \qquad i,j=1,2,...,2n+2  \label{re.13}
\end{equation}

Substituting (\ref{re.1}) and (\ref{re.12}) into (\ref{int.3}), we have in
both $16(n+1)^{4}$ functional equations for the $k_{ij}$ elements, many of
which are dependent. In order to solve them, we shall proceed in the
following way. First we consider the $(i,j)$ component of the matrix
equation (\ref{int.3}). By differentiating it with respect to $v$ and taking 
$v=0$, we will get algebraic equations involving the single variable $u$ and 
$4(n+1)^{2}$ parameters 
\begin{equation}
\beta _{ij}=\frac{dk_{ij}(v)}{dv}|_{v=0}\qquad i,j=1,2,...,2n+2
\label{re.14}
\end{equation}%
Second, these algebraic equations are denoted by $E[i,j]=0$ and collected
into blocks $B[i,j]$ , $i=1,...,2(n+1)^{2}$ and $j=i,i+1,...,(2n+2)^{2}-i$,
defined by 
\begin{equation}
B[i,j]=\left\{ 
\begin{array}{c}
E[i,j]=0,\ E[j,i]=0,\  \\ 
E[4(n+1)^{2}+1-i,4(n+1)^{2}+1-j]=0, \\ 
E[4(n+1)^{2}+1-j,4(n+1)^{2}+1-i]=0.%
\end{array}%
\right. \   \label{re.15}
\end{equation}%
For a given block $B[i,j]$, the equation $E[4(n+1)^{2}+1-i,4(n+1)^{2}+1-j]=0$
can be obtained from the equation $E[i,j]=0$ by interchanging 
\begin{eqnarray}
k_{ij} &\longleftrightarrow &k_{i^{\prime }j^{\prime }},\quad \beta
_{ij}\longleftrightarrow \beta _{i^{\prime }j^{\prime }},\quad \ \
b_{i}^{\pm }\longleftrightarrow b_{i^{\prime }}^{\pm }\ ,\quad
a_{ij}\leftrightarrow a_{i^{\prime }j^{\prime },}  \nonumber \\
a_{3} &\longleftrightarrow &a_{4},\quad \ \ a_{6}\longleftrightarrow
a_{6},\quad \quad \ \ a_{7}\longleftrightarrow a_{8}\ .  \label{re.16}
\end{eqnarray}%
and the equation $E[j,i]=0$ is obtained from the equation $E[i,j]=0$ by the
interchanging 
\begin{equation}
k_{ij}\longleftrightarrow k_{ji},\quad \beta _{ij}\longleftrightarrow \beta
_{ji},\quad a_{ij}\longleftrightarrow a_{j^{\prime }i^{\prime }}\ .\quad 
\label{re.17}
\end{equation}%
In this way, we can control all equations and a particular solution is
simultaneously connected with at least four equations.

Since the $R$-matrix (\ref{re.1}) satisfies unitarity, P and T invariances
and crossing symmetry, the matrix $K_{+}(u)$\ is obtained using (\ref{int.4}%
) with the following $M$-matrix%
\begin{equation}
M_{ij}=\delta _{ij\ }q^{2n+3-2\overset{\_}{i}},\qquad i,j=1,2,...,2n+2
\label{re.19}
\end{equation}

\section{The $D_{n+1}^{(2)}$ K-matrix solutions}

Analyzing the $D_{n+1}^{(2)}$ {\small RE }\ one can see that they possess a
special structure. A lot of equations exist involving only the elements out
of \ a block diagonal structure which consist of the diagonal elements $k_{ii%
\text{ }}$plus the central elements of the secondary diagonal, $k_{n+1,n+2}$
and $k_{n+2,n+1}$. This block diagonal structure commutes with $n$ distinct $%
U(1)$ symmetries, the minimal symmetry of the $R$-matrix \cite{MG}. Through
this structure we can identify two types of solutions: the type-I solution
which is a general $K$-matrix solution with all matrix elements non-null and
the type-II solution for which the $K$-matrix has this block diagonal
structure. Moreover, for the models with a even number of $U(1)$ conserved
charge, we can find a third type of solution which manifests the $%
U(1)\otimes U(1)$ symmetries. Therefore for $n$ even we can find the
type-III solution which is a diagonal \ $K$-matrix. These solutions are to
each other unyielding.

Having identified these possibilities we may proceed in order to find
explicitly the regular solutions. We start looking for the type-I solution.

\subsection{The type-I solution}

The simplest {\small RE} are those involving only the elements of the
secondary diagonal. We chose to express their solutions in terms of the
element $k_{1,2n+2}$: 
\begin{equation}
k_{ii^{\prime }}=\frac{\beta _{ii^{\prime }}}{\beta _{1,2n+2}}%
k_{1,2n+2},\qquad i\neq \{n+1,n+2\}.  \label{str.1}
\end{equation}%
From the collections $\{B[i,j]\}$, $i=1,2,...,n-1$ one can see that the 
{\small RE} of the last blocks of each collection are simple and we can
easily solve them expressing the elements $k_{ij}$ with $j\neq i^{\prime }$
in function of $k_{1,2n+2}$ 
\begin{eqnarray}
k_{ij} &=&\left\{ 
\begin{array}{c}
\digamma _{ij}\left( \beta _{ij}a_{3}a_{ii}-\beta _{j^{\prime }i^{\prime
}}a_{2}a_{ij^{\prime }}\right) k_{1,2n+2},\qquad (i<j^{\prime }) \\ 
\\ 
\digamma _{ij}\left( \beta _{ij}a_{4}a_{ii}-\beta _{j^{\prime }i^{\prime
}}a_{2}a_{ij^{\prime }}\right) k_{1,2n+2},\qquad (i>j^{\prime })%
\end{array}%
\right.   \nonumber \\
&&  \label{str.2}
\end{eqnarray}%
where 
\begin{equation}
\digamma _{ij}=\frac{a_{1}a_{ii}-a_{2}^{2}}{\beta _{1,2n+2}\left(
a_{ii}^{2}a_{3}a_{4}-a_{2}^{2}a_{ij^{\prime }}a_{j^{\prime }i}\right) }.
\label{str.3}
\end{equation}%
Moreover, for $j=n+1,n+2$ with $i\neq j,j^{\prime }$ we have 
\begin{eqnarray}
k_{ij} &=&\left\{ 
\begin{array}{c}
\Delta _{i}\left[ a_{ii}\left( \beta _{ij}a_{5}+\beta _{ij^{\prime
}}a_{7}\right) -a_{2}\left( \beta _{j^{\prime }i^{\prime }}b_{i}^{+}+\beta
_{ji^{\prime }}b_{i}^{-}\right) \right] k_{1,2n+2},\qquad (i^{\prime }>j) \\ 
\\ 
\Delta _{i}\left[ a_{ii}\left( \beta _{ij}a_{6}+\beta _{ij^{\prime
}}a_{8}\right) -a_{2}\left( \beta _{j^{\prime }i^{\prime }}b_{i}^{+}+\beta
_{ji^{\prime }}b_{i}^{-}\right) \right] k_{1,2n+2},\qquad (i^{\prime }<j)%
\end{array}%
\right.   \nonumber \\
&&  \label{str.4}
\end{eqnarray}%
and for $i=n+1,n+2$ with $j\neq i,i^{\prime }$ we get 
\begin{eqnarray}
k_{ij} &=&\left\{ 
\begin{array}{c}
\Delta _{j}\left[ a_{jj}\left( \beta _{ij}a_{5}+\beta _{i^{\prime
}j}a_{7}\right) -a_{2}\left( \beta _{j^{\prime }i^{\prime }}b_{j}^{+}+\beta
_{j^{\prime }i}b_{j}^{-}\right) \right] k_{1,2n+2},\qquad (i^{\prime }>j) \\ 
\\ 
\Delta _{j}\left[ a_{jj}\left( \beta _{ij}a_{6}+\beta _{i^{\prime
}j}a_{8}\right) -a_{2}\left( \beta _{j^{\prime }i^{\prime }}b_{j}^{+}+\beta
_{j^{\prime }i}b_{j}^{-}\right) \right] k_{1,2n+2},\qquad (i^{\prime }<j)%
\end{array}%
\right.   \nonumber \\
&&  \label{str.5}
\end{eqnarray}%
where 
\begin{equation}
\Delta _{l}=\frac{a_{1}a_{ll}-a_{2}^{2}}{\beta _{1,2n+2}\left(
a_{ll}^{2}(a_{6}+a_{8})(a_{5}+a_{7})-a_{2}^{2}(b_{l}^{+}+b_{l}^{-})(b_{l^{%
\prime }}^{+}+b_{l^{\prime }}^{-})\right) }  \label{str.6}
\end{equation}%
Here we observe that for the $D_{2}^{(2)}$ model, \ $\digamma _{ij}=0$ and $%
\Delta _{l}=\frac{0}{0}$. However, through an appropriate choice of $\Delta
_{l}$ we can include the case $n=1$ in our discussion (see appendix A).

Substituting these expressions in the remaining {\small RE \ ( }in fact, it
is enough to consider the equations of the collections $\{B[1,j]\}$ and $%
\{B[2,j]\}$ ), and looking at the equations of the type 
\begin{equation}
G(u)k_{1,2n+2}(u)=0  \label{str.7}
\end{equation}%
where $G(u)=\sum_{k}f_{k}(\left\{ \beta _{ij}\right\} ){\rm e}^{ku}$. \ The
constraint equations $f_{k}(\left\{ \beta _{ij}\right\} )\equiv 0,\forall k$%
, can be solved in terms of the $2n+2$ parameters which allow us to find all 
$k_{ij}$ elements out of the block-diagonal structure in terms of the $%
k_{1,2n+2}$.

Of course the expressions for $k_{ij}$ will depend on our choice of these
parameters. After some attempts we concluded the choice $\beta _{12}$, $%
\beta _{13}$, ..., $\beta _{1,2n+2}$ and $\beta _{21}$ as the most
appropriate for our purpose.

Taking into account the fixed\ parameters and the Boltzmann weights defined
above we can rewrite these $k_{ij}$ matrix elements for $n\neq 1$ in the
following way:

The elements in the secondary diagonal of $K_{-}(u)$ are given by 
\begin{equation}
k_{i,i^{\prime }}(u)=q^{\overset{\_}{i}-2n}\left( \frac{q^{n-1}+1}{q+1}%
\right) ^{2}\left( \frac{\beta _{1,i^{\prime }}}{\beta _{1,2n+2}}\right)
^{2}k_{1,2n+2}(u),\qquad (i\neq 1,2n+2)  \label{str.8}
\end{equation}%
and 
\begin{equation}
k_{2n+2,1}(u)=q^{2n-3}\left( \frac{\beta _{21}}{\beta _{1,2n+1}}\right)
^{2}k_{1,2n+2}(u).  \label{str.9}
\end{equation}%
The elements of the first row and the elements of the first column are,
respectively given by%
\begin{equation}
k_{1,j}(u)=\left( \frac{q^{n-1}+1}{{\rm e}^{2u}+q^{n-1}}\right) \left( \frac{%
\Gamma _{1,j}(u)}{\beta _{1,2n+2}}\right) k_{1,2n+2}(u),\qquad (j\neq 2n+2)
\label{str.10a}
\end{equation}%
\begin{equation}
k_{i,1}(u)=q^{\overset{\_}{i}-3}\left( \frac{q^{n-1}+1}{{\rm e}^{2u}+q^{n-1}}%
\right) \left( \frac{\Gamma _{1,i^{\prime }}(u)}{\beta _{1,2n+2}}\right)
\left( \frac{\beta _{21}}{\beta _{1,2n+1}}\right) k_{1,2n+2}(u),\qquad
(i\neq 2n+2)  \label{str.10b}
\end{equation}%
while for the last column and for the last row, we have%
\begin{equation}
k_{i,2n+2}(u)=q^{\overset{\_}{i}-n-2}\left( \frac{q^{n-1}+1}{{\rm e}%
^{2u}+q^{n-1}}\right) \left( \frac{\Pi _{1,i^{\prime }}(u)}{\beta _{1,2n+2}}%
\right) {\rm e}^{2u}k_{1,2n+2}(u),\qquad (i\neq 1)  \label{str.11a}
\end{equation}%
\begin{equation}
k_{2n+2,j}(u)=q^{n-2}\left( \frac{q^{n-1}+1}{{\rm e}^{2u}+q^{n-1}}\right)
\left( \frac{\beta _{21}}{\beta _{1,2n+1}}\right) \left( \frac{\Pi _{1,j}(u)%
}{\beta _{1,2n+2}}\right) {\rm e}^{2u}k_{1,2n+2}(u),\qquad (j\neq 1)
\label{str.11b}
\end{equation}%
The remaining matrix elements are given by%
\begin{equation}
k_{ij}(u)=q^{\overset{\_}{i}-n-1}\left( \frac{q^{n-1}+1}{q+1}\right) \left( 
\frac{q^{n-1}+1}{{\rm e}^{2u}+q^{n-1}}\right) \left( \frac{\Gamma
_{1,i^{\prime }}(u)}{\beta _{1,2n+2}}\right) \left( \frac{\Gamma _{1,j}(u)}{%
\beta _{1,2n+2}}\right) k_{1,2n+2}(u),  \label{str.12}
\end{equation}%
for $i^{\prime }>j$ and%
\begin{equation}
k_{ij}(u)=q^{\overset{\_}{i}-2n-1}\left( \frac{q^{n-1}+1}{q+1}\right) \left( 
\frac{q^{n-1}+1}{{\rm e}^{2u}+q^{n-1}}\right) \left( \frac{\Pi _{1,i^{\prime
}}(u)}{\beta _{1,2n+2}}\right) \left( \frac{\Pi _{1,j}(u)}{\beta _{1,2n+2}}%
\right) {\rm e}^{2u}k_{1,2n+2}(u)  \label{str.13}
\end{equation}%
for $i^{\prime }<j$. \ 

In these expressions we are using a compact notation defined by

\begin{equation}
\Gamma _{1,a}(u)=\left\{ 
\begin{array}{c}
\beta _{1,a},\qquad \qquad \quad \quad (a\neq n+1,n+2) \\ 
\\ 
\frac{1}{2}\left( e^{u}\beta _{-}+\beta _{+}\right) ,\qquad \quad \quad
(a=n+1) \\ 
\\ 
\frac{1}{2}\left( -e^{u}\beta _{-}+\beta _{+}\right) ,\qquad \quad \ (a=n+2)%
\end{array}%
\right. \qquad   \label{str.14}
\end{equation}%
and%
\begin{equation}
\Pi _{1,a}(u)=\left\{ 
\begin{array}{c}
\beta _{1,a},\quad \qquad \qquad \quad \quad (a\neq n+1,n+2) \\ 
\\ 
\frac{1}{2}\left( q^{n}e^{-u}\beta _{-}+\beta _{+}\right) ,\quad \qquad
\quad (a=n+1) \\ 
\\ 
\frac{1}{2}\left( -q^{n}e^{-u}\beta _{-}+\beta _{+}\right) ,\quad \qquad \
(a=n+2)%
\end{array}%
\right.   \label{str.15}
\end{equation}%
where $\beta _{\pm }=\beta _{1,n+1}\pm \beta _{1,n+2}$. \ To include the
case $n=1$ we will make some modifications in these expressions (see
appendix A).

At this point we find the $2n(2n+3)$ matrix elements in terms of the $2n+2$
parameters. However, we still need to find the $2n+4$ matrix elements that
belong to the block diagonal.

The block diagonal structure has the form%
\begin{equation}
{\rm diag}\left( k_{11},k_{22},\cdots ,k_{nn},{\cal B},k_{n+3,n+3},\cdots
,k_{2n+2,2n+2}\right) ,  \label{str.16a}
\end{equation}%
$\ $where ${\cal B}$ \ contains the central elements 
\begin{equation}
{\cal B}=\left( 
\begin{array}{c}
\begin{array}{cc}
k_{n+1,n+1} & k_{n+1,n+2} \\ 
k_{n+2,n+1} & k_{n+2,n+2}%
\end{array}%
\end{array}%
\right)   \label{srt.16b}
\end{equation}%
Here the situation is a little bit different. It is very cumbersome to write
these matrix elements in terms of the Boltzmann weights. But, after some
algebraic manipulations, it is possible to see that the diagonal elements
satisfy two distinct recurrence relations%
\begin{equation}
k_{ii}=\left\{ 
\begin{array}{c}
k_{11}-\left( \frac{q^{n-1}+1}{{\rm e}^{2u}+q^{n-1}}\right) \left( \frac{%
\beta _{11}-\beta _{ii}}{\beta _{1,2n+2}}\right) k_{1,2n+2},\hfill {}\qquad
(i<n+1) \\ 
\\ 
k_{n+3,n+3}-\left( \frac{q^{n-1}+1}{{\rm e}^{2u}+q^{n-1}}\right) \left( 
\frac{\beta _{n+3,n+3}-\beta _{ii}}{\beta _{1,2n+2}}\right) {\rm e}%
^{2u}k_{1,2n+2},\hfill \qquad (i>n+2)%
\end{array}%
\right.   \label{str.17}
\end{equation}

Substituting (\ref{str.17}) into the {\small RE} we can find $k_{11}$ and $%
k_{n+3,n+3}$ and consequently, all $k_{ii}$ $\notin {\cal B}$ will be known
after finding the $2n$ parameters $\beta _{ii}$.

The solution of this problem depends on the parity of $n$. Besides, in this
stage, all remaining parameters $\beta _{ij}$, including those associated
with the central elements, are fixed in terms of $n+3$ parameters:

For $n$ odd the solution is

\begin{eqnarray}
k_{11}(u) &=&\left\{ \left( \frac{q^{n-1}+1}{q+1}\right) \frac{(q+1)({\rm e}%
^{2u}+1)(q^{n}\beta _{-}^{2}-\beta _{+}^{2})+2({\rm e}^{2u}-q^{n})(q^{n}%
\beta _{-}^{2}+\beta _{+}^{2})}{8\beta _{1,2n+2}^{2}\ q^{n-1/2}({\rm e}%
^{2u}+1)}\right.  \nonumber \\
&&\left. +\frac{2q(q^{n-1}-1)+(q-1)({\rm e}^{2u}+1)}{\beta
_{1,2n+2}(q^{n}-1)({\rm e}^{2u}-1)}\right\} \left( \frac{q^{n-1}+1}{{\rm e}%
^{2u}+q^{n-1}}\right) k_{1,2n+2}(u)  \label{odd.1}
\end{eqnarray}%
\begin{eqnarray}
k_{n+3,n+3}(u)\!\! &=&\!\!\left\{ \left( \frac{q^{n-1}+1}{q+1}\right) \frac{%
(q+1)({\rm e}^{2u}+1)(q^{n}\beta _{-}^{2}-\beta _{+}^{2})-2({\rm e}%
^{2u}-q^{n})(q^{n}\beta _{-}^{2}+\beta _{+}^{2})}{8\beta _{1,2n+2}^{2}\
q^{n-1/2}({\rm e}^{2u}+1)}\right.  \nonumber \\
&&\left. +\frac{2q(q^{n-1}-1)+(q-1)({\rm e}^{2u}+1)}{\beta
_{1,2n+2}(q^{n}-1)({\rm e}^{2u}-1)}\right\} \left( \frac{q^{n-1}+1}{{\rm e}%
^{2u}+q^{n-1}}\right) {\rm e}^{2u}k_{1,2n+2}(u)  \nonumber \\
&&  \label{odd.2}
\end{eqnarray}

The parameters $\beta _{ii}$ , $i\neq n+1,n+2$ are fixed by the following
recurrence relations 
\begin{equation}
\beta _{i+1,i+1}=\left\{ 
\begin{array}{c}
\beta _{ii}+(-q)^{i-1}\Theta _{{\rm odd}}\qquad \qquad (i<n) \\ 
\\ 
\beta _{ii}+(-q)^{i-n-3}\Theta _{{\rm odd}}\qquad \qquad (i>n+2)%
\end{array}%
\right.  \label{odd.3}
\end{equation}%
with 
\begin{equation}
\beta _{n+3,n+3}=\beta _{11}+2+\left( \frac{q^{n-1}+1}{q+1}\right) \frac{%
(q^{n}-1)(q^{n}\beta _{-}^{2}+\beta _{+}^{2})}{4q^{n-1/2}\beta _{1,2n+2}}
\label{odd.4}
\end{equation}%
and 
\begin{equation}
\Theta _{{\rm odd}}=-\frac{(q+1)^{2}}{q^{n}-1}-\frac{(q+1)(q^{n-1}+1)(\beta
_{-}^{2}q^{n}-\beta _{+}^{2})}{8q^{n-1/2}\beta _{1,2n+2}}  \label{odd.5}
\end{equation}

\bigskip Finally, we can solve the last {\small RE} to find the central
elements. The solution is%
\begin{eqnarray}
k_{n+1,n+1}(u) &=&-\left\{ \left( \frac{q^{n-1}+1}{q+1}\right) \frac{%
q^{n}\beta _{-}^{2}-\beta _{+}^{2}}{8q^{n-1/2}}({\rm e}^{2u}+q^{n})\right.  
\nonumber \\
&&\left. +\beta _{1,2n+2}\frac{({\rm e}^{2u}+1)({\rm e}^{2u}-q^{n})}{({\rm e}%
^{2u}-1)(q^{n}-1)}\right\} \left( \frac{q^{n-1}+1}{{\rm e}^{2u}+q^{n-1}}%
\right) \frac{k_{1,2n+2}(u)}{\beta _{1.2n+2}^{2}},  \nonumber \\
k_{n+2,n+2}(u) &=&k_{n+1,n+1}(u)  \label{odd.6}
\end{eqnarray}%
\begin{eqnarray}
k_{n+1,n+2}(u)\! &=&\!\frac{(q^{n-1}+1)^{2}\left[ (q^{n}+1)(q^{n}\beta
_{-}^{2}+\beta _{+}^{2}){\rm e}^{u}-2q^{n}\beta _{-}\beta _{+}({\rm e}%
^{2u}+1)\right] }{4q^{n-1/2}(q+1)({\rm e}^{2u}+1)({\rm e}^{2u}+q^{n-1})}%
\frac{{\rm e}^{u}k_{1,2n+2}(u)}{\beta _{1,2n+2}^{2}},  \nonumber \\
&&  \label{odd.7a}
\end{eqnarray}%
\begin{eqnarray}
k_{n+2,n+1}(u)\! &=&\!\frac{(q^{n-1}+1)^{2}\left[ (q^{n}+1)(q^{n}\beta
_{-}^{2}+\beta _{+}^{2}){\rm e}^{u}+2q^{n}\beta _{-}\beta _{+}({\rm e}%
^{2u}+1)\right] }{4q^{n-1/2}(q+1)({\rm e}^{2u}+1)({\rm e}^{2u}+q^{n-1})}%
\frac{{\rm e}^{u}k_{1,2n+2}(u)}{\beta _{1,2n+2}^{2}}.  \nonumber \\
&&  \label{odd.7b}
\end{eqnarray}%
Moreover, there are $n$ parameters which have been fixed by the {\small RE}

\begin{equation}
\beta _{21}=\frac{1}{q^{2n-3}}\left( \frac{q^{n-1}+1}{q+1}\right) ^{2}\frac{%
\beta _{1,n}\beta _{1,n+3}\beta _{1,2n+1}}{\beta _{1,2n+2}^{2}},\qquad
(n\neq 1)  \label{odd.8}
\end{equation}%
\begin{equation}
\beta _{1,n}=\frac{\left[ (q^{n}-1)(q^{n-1}+1)(q^{n}\beta _{-}^{2}-\beta
_{+}^{2})+8q^{n-1/2}(q+1)\beta _{1,2n+2}\right] (q+1)}{8\sqrt{q}%
(q^{n}-1)(q^{n-1}+1)\beta _{1,n+3}},\qquad (n\neq 1)  \label{odd.9}
\end{equation}%
\begin{equation}
\beta _{1j}=(-)^{j-1}\frac{\beta _{1,n}\beta _{1,n+3}}{\beta _{1,2n+3-j}}%
,\qquad j=2,3,\cdots ,n-1.  \label{odd.10}
\end{equation}%
The final result is a general solution with $n+3$ free parameters $\beta
_{11}$, $\beta _{1,n+1}$, $\beta _{1,n+2}$, ..., $\beta _{1,2n+2}$.

By the choose 
\begin{equation}
k_{1,2n+2}(u)=\frac{1}{2}\beta _{1,2n+2}\left( {\rm e}^{2u}-1\right) ,
\label{odd.11}
\end{equation}%
one can , for instance, to fix the parameter $\beta _{11}$ using the regular
condition (\ref{re.13}) and in that way to end with a $\ (n+2)$-parameter
solution for $D_{n+1}^{(2)}$ models with $n$ odd.

The corresponding matrix $K_{+}(u)$ \ is obtained used (\ref{int.4}) \ and
in this case we have {\rm Tr}$(K_{+}(0))\neq 0$ . Therefore, the equation (%
\ref{int.2}) gives the corresponding integrable open chain Hamiltonians,
where $H_{k,k+1}$ are the tensors $s_{k,k+1}(q)$ derived by Jimbo \cite%
{Jimbo}.

Now we turn to describe the even solution. \ In the $n$-even case we find
the following expressions for $k_{11}$ and $k_{n+3,n+3}$, respectively 
\begin{eqnarray}
k_{11}(u) &=&\frac{k_{1,2n+2}(u)}{2q^{n-1/2}\beta _{1,2n+2}^{2}({\rm e}%
^{4u}-1)({\rm e}^{2u}+q^{n-1})}\left\{ (q+1)({\rm e}^{2u}-q^{n})(q^{n}\beta
_{-}^{2}{\rm e}^{2u}-\beta _{+}^{2})\right.  \nonumber \\
&&\left. +2\sqrt{q}\ \beta _{1,n}\beta _{1,n+3}({\rm e}^{2u}-1)\left[ 2({\rm %
e}^{2u}-q^{n})-(q+1)({\rm e}^{2u}+1)\right] \right\}  \nonumber \\
&&  \label{even.1}
\end{eqnarray}%
\begin{eqnarray}
k_{n+3,n+3}(u) &=&\frac{{\rm e}^{2u}k_{1,2n+2}(u)}{2q^{n-1/2}\beta
_{1,2n+2}^{2}({\rm e}^{4u}-1)({\rm e}^{2u}+q^{n-1})}\left\{ (q+1)({\rm e}%
^{2u}-q^{n})(q^{n}\beta _{-}^{2}-\beta _{+}^{2}{\rm e}^{2u})\right. 
\nonumber \\
&&\left. -2\sqrt{q}\ \beta _{1,n}\beta _{1,n+3}({\rm e}^{2u}-1)\left[ 2({\rm %
e}^{2u}-q^{n})-(q+1)({\rm e}^{2u}+1)\right] \right\}  \nonumber \\
&&  \label{even.2}
\end{eqnarray}%
The central elements are given by%
\begin{equation}
k_{n+1,n+1}(u)=k_{n+2,n+2}(u)=2\left( \frac{{\rm e}^{2u}}{{\rm e}^{2u}-1}%
\right) \left( \frac{q^{n-1}+1}{{\rm e}^{2u}+q^{n-1}}\right) \frac{%
k_{1,2n+2}(u)}{\beta _{1,2n+2}}  \label{even.3}
\end{equation}%
\begin{eqnarray}
k_{n+1,n+2}(u)\!\! &=&\!\!\frac{k_{1,2n+2}(u)}{4q^{n-1/2}\beta _{1,2n+2}^{2}(%
{\rm e}^{2u}+q^{n-1})({\rm e}^{2u}+1)}\!\left( \frac{q^{n-1}+1}{q+1}\right)
^{2}  \nonumber \\
&&\left\{ \left( q^{n}\beta _{-}^{2}+\beta _{+}^{2}\right) ^{2}(q+1)(q^{n}+1)%
{\rm e}^{2u}-2\beta _{-}\beta _{+}(q+1)q^{n}{\rm e}^{u}({\rm e}%
^{2u}+1)\right. \!  \nonumber \\
&&\left. -4\sqrt{q}\beta _{1,n}\beta _{1,n+3}({\rm e}^{2u}-q^{n})({\rm e}%
^{2u}-1)\right\}  \label{even.4a}
\end{eqnarray}%
\begin{eqnarray}
k_{n+2,n+1}(u)\!\! &=&\!\!\frac{k_{1,2n+2}(u)}{4q^{n-1/2}\beta _{1,2n+2}^{2}(%
{\rm e}^{2u}+q^{n-1})({\rm e}^{2u}+1)}\left( \frac{q^{n-1}+1}{q+1}\right)
^{2}\!  \nonumber \\
&&\left\{ \!\left( q^{n}\beta _{-}^{2}+\beta _{+}^{2}\right)
^{2}(q+1)(q^{n}+1){\rm e}^{2u}+2\beta _{-}\beta _{+}(q+1)q^{n}{\rm e}^{u}(%
{\rm e}^{2u}+1)\right.  \nonumber \\
&&\left. -4\sqrt{q}\beta _{1,n}\beta _{1,n+3}({\rm e}^{2u}-q^{n})({\rm e}%
^{2u}-1)\right\}  \label{even;4b}
\end{eqnarray}%
The $\beta _{ii}$ parameters are fixed by the following recurrence relations

\begin{equation}
\beta _{i+1,i+1}=\left\{ 
\begin{array}{c}
\beta _{ii}+(-q)^{i-1}\Theta _{{\rm even}},\qquad \qquad (i<n) \\ 
\\ 
\beta _{ii}-(-q)^{i-n-3}\Theta _{{\rm even}},\qquad \qquad (i>n+2)%
\end{array}%
\right.   \label{even.5}
\end{equation}%
with 
\begin{equation}
\beta _{n+3,n+3}=\beta _{11}+2-\frac{2(q^{n}-1)(q+1)(q^{n}\beta
_{-}^{2}+\beta _{+}^{2})+8\beta _{1,n}\beta _{1,n+3}q^{3/2}(q^{n-1}+1)}{%
(q+1)(q^{n}-1)(q^{n}\beta _{-}^{2}-\beta _{+}^{2})}  \label{even.6}
\end{equation}%
and 
\begin{equation}
\Theta _{{\rm even}}=-8\frac{\sqrt{q}(q+1)}{q^{n}-1}\frac{\beta _{1,n}\beta
_{1,n+3}}{\beta _{-}^{2}q^{n}-\beta _{+}^{2}}  \label{even.7}
\end{equation}%
Now the $n$ fixed parameters are%
\begin{equation}
\beta _{21}=-q^{3-2n}\left( \frac{q^{n-1}+1}{q+1}\right) \beta _{1,2n+1}%
\frac{\beta _{1,n}\beta _{1,n+3}}{\beta _{1,2n+2}^{2}}  \label{even.8a}
\end{equation}%
\begin{equation}
\beta _{1,2n+2}=-\frac{1}{8}\left( \frac{q^{n}-1}{q^{n-1/2}}\right) \left( 
\frac{q^{n-1}+1}{q+1}\right) (q^{n}\beta _{-}^{2}-\beta _{+}^{2})
\label{even.8b}
\end{equation}

\begin{equation}
\beta _{1,j}=(-)^{j-1}\frac{\beta _{1,n}\beta _{1,n+3}}{\beta _{1,2n+3-j}}%
,\qquad j=2,3,\cdots ,n-1  \label{even.8c}
\end{equation}%
and the $n+3$ free parameters are $\beta _{11}$,$\beta _{1,n}$,..., $\beta
_{1,2n+1}$. 

Again, one can fix $\beta _{11}$ from the regular condition to get a general
solution with $n+2$ free parameters. Here we also have {\rm Tr}$%
(K_{+}(0))\neq 0$, $\forall n$ \ and (\ref{int.2}) gives the corresponding
integrable open chain Hamiltonians.

There are many particular solutions which are obtained by vanishing some of
these free parameters. Despite the existence of $n+2$ free parameters in a
type-I solution \ it seems impossible to make a reduction through the
annulment of certain parameters in order to obtain a type-II or type-III
solution.

\subsection{The Type-II solution}

As was already mentioned, the $D_{n+1}^{(2)}$ models have $n$ distinct $U(1)$
conserved charges, and the $K$-matrix ansatz compatible with these
symmetries is the block diagonal structure presented in the previous section.

Looking for the general solution of the corresponding {\small RE} \ we find
that the only possible solution is obtained when the two recurrence
relations (\ref{str.17}) are degenerated into $k_{11}$ and into $k_{n+3,n+3}$%
, respectively.

\begin{equation}
k_{n,n}(u)=k_{n-1,n-1}(u)=\cdots =k_{22}(u)=k_{11}(u),  \label{bd.1a}
\end{equation}%
and%
\begin{equation}
k_{2n+2,2n+2}(u)=k_{2n+1,2n+1}(u)=\cdots =k_{n+4,n+4}(u)=k_{n+3,n+3}(u)
\label{bd.1b}
\end{equation}
The  type-II solution can be obtained by the same procedure described before
and in what follows we only quote the final results.

We have found two solutions for any value of $n$. \ The first solution is
given by%
\begin{equation}
k_{11}(u)=\frac{1}{2}\frac{({\rm e}^{2u}+q^{n})\left\lfloor (q^{n}-1)({\rm e}%
^{2u}+1)-\beta _{n+1,n+2}(q^{n}+1)({\rm e}^{2u}-1)\right\rfloor }{{\rm e}%
^{2u}(q^{2n}-1)}  \label{bd.2}
\end{equation}%
\begin{equation}
k_{n+3,n+3}(u)=\frac{1}{2}\frac{({\rm e}^{2u}+q^{n})\left\lfloor (q^{n}-1)(%
{\rm e}^{2u}+1)+\beta _{n+1,n+2}(q^{n}+1)({\rm e}^{2u}-1)\right\rfloor }{%
(q^{2n}-1)}  \label{bd.2b}
\end{equation}%
with central elements 
\begin{equation}
k_{n+1,n+2}(u)=k_{n+2,n+1}(u)=\frac{1}{2}\beta _{n+1,n+2}({\rm e}^{2u}-1)
\label{bd.3}
\end{equation}%
\begin{equation}
k_{n+1,n+1}(u)=\frac{1}{2}({\rm e}^{2u}+1)\left\{ 1+\frac{({\rm e}^{2u}-1)}{%
{\rm e}^{u}(q^{2n}-1)}\Gamma _{\pm }\right\}   \label{bd.4}
\end{equation}%
\begin{equation}
k_{n+2,n+2}(u)=\frac{1}{2}({\rm e}^{2u}+1)\left\{ 1+\frac{({\rm e}^{2u}-1)}{%
{\rm e}^{u}(q^{2n}-1)}\Gamma _{\mp }\right\}   \label{bd.4b}
\end{equation}%
where%
\begin{eqnarray}
\Gamma _{\pm } &=&\frac{1}{\Sigma _{\pm }}\left\{ 2q^{n}\left[
(q^{n}+1)^{2}\beta _{n+1,n+2}^{2}-(q^{n}-1)^{2}\right] \right.   \nonumber \\
&&\pm \left. \left[ (q^{n}+1)^{2}\beta _{n+1,n+2}+(q^{n}-1)^{2}\right] \sqrt{%
q^{n}\left[ (q^{n}+1)^{2}\beta _{n+1,n+2}^{2}-(q^{n}-1)^{2}\right] }\right\} 
\nonumber \\
&&  \label{bd.5}
\end{eqnarray}%
and%
\begin{equation}
\Sigma _{\pm }=\left[ (q^{n}+1)^{2}\beta _{n+1,n+2}+(q^{n}-1)^{2}\right] \pm
2\sqrt{q^{n}\left[ (q^{n}+1)^{2}\beta _{n+1,n+2}^{2}-(q^{n}-1)^{2}\right] }
\label{bd.6}
\end{equation}%
The signs ($\pm $) and ($\pm $) are indicating the existence of two
conjugated solutions.  Here we notice that these solutions degenerated into
two complex diagonal solutions when $\beta _{n+1,n+2}=0.$ 

For the second solution we have%
\begin{eqnarray}
k_{11}(u) &=&\frac{1}{2}\frac{({\rm e}^{2u}-q^{n})}{{\rm e}^{u}(q^{n}-1)^{2}}%
\left\{ ({\rm e}^{2u}-1)\left[ (q^{n}+1)\beta _{n+1,n+2}\pm 2\sqrt{q^{n}%
\left[ \beta _{n+1,n+2}^{2}-1\right] }\right] \right.  \nonumber \\
&&-\left. ({\rm e}^{2u}+1)(q^{n}-1)\right\}  \label{bd.7}
\end{eqnarray}

\begin{eqnarray}
k_{n+3,n+3}(u) &=&-\frac{1}{2}\frac{{\rm e}^{u}({\rm e}^{2u}-q^{n})}{%
(q^{n}-1)^{2}}\left\{ ({\rm e}^{2u}-1)\left[ (q^{n}+1)\beta _{n+1,n+2}\pm 2%
\sqrt{q^{n}\left[ \beta _{n+1,n+2}^{2}-1\right] }\right] \right.   \nonumber
\\
&&+\left. ({\rm e}^{2u}+1)(q^{n}-1)\right\}   \label{bd.7b}
\end{eqnarray}%
with the following central elements%
\begin{equation}
k_{n+1,n+1}(u)=k_{n+2,n+2}(u)=\frac{1}{2}{\rm e}^{u}({\rm e}^{2u}+1)
\label{bd.9}
\end{equation}%
\begin{eqnarray}
k_{n+1,n+2}(u) &=&\frac{1}{2}\frac{({\rm e}^{2u}-1)}{(q^{n}-1)^{2}}\left\{
\beta _{n+1,n+2}\left[ {\rm e}^{u}(q^{n}+1)^{2}-2q^{n}({\rm e}^{2u}+1)\right]
\right.   \nonumber \\
&&\mp \left. (q^{n}+1)\sqrt{q^{n}\left[ \beta _{n+1,n+2}^{2}-1\right] }({\rm %
e}^{u}-1)^{2}\right\}   \label{bd.9c}
\end{eqnarray}%
\begin{eqnarray}
k_{n+2,n+1}(u) &=&\frac{1}{2}\frac{({\rm e}^{2u}-1)}{(q^{n}-1)^{2}}\left\{
\beta _{n+1,n+2}\left[ {\rm e}^{u}(q^{n}+1)^{2}+2q^{n}({\rm e}^{2u}+1)\right]
\right.   \nonumber \\
&&\pm \left. (q^{n}+1)\sqrt{q^{n}\left[ \beta _{n+1,n+2}^{2}-1\right] }({\rm %
e}^{u}+1)^{2}\right\}   \label{bd.9b}
\end{eqnarray}%
Unlike of the first solution does not exist a way of deriving a diagonal
solution starting from these conjugated solutions. Moreover, when $n=1$\ we
have ${\rm Tr}(K_{+}(0))=0$. Therefore in this case the corresponding
boundary term in the integrable open chain Hamiltonian is not more given by (%
\ref{int.2}) but taking into account the second order expansion of the
transfer matrix in the spectral parameter $u$ \cite{Cuerno}.

\subsection{The Type-III Solution}

For the models with an even number of $U(1)$ conserved charge, we look for
diagonal $K$-matrix solutions which manifest the $U(1)\otimes U(1)$
symmetries. \ Here there is only one solution, namely the ''almost unity''
solution \cite{MG} 
\begin{eqnarray}
k_{11}(u) &=&{\rm e}^{-2u}  \nonumber \\
k_{22}(u) &=&k_{33}(u)=\cdots =k_{2n+1,2n+1}(u)=1  \nonumber \\
k_{2n+2,2n+2}(u) &=&{\rm e}^{2u}  \label{dd.1}
\end{eqnarray}%
In that point we finished noticing the absence of the trivial solution $%
K_{-}(u)=1$ for this system.

The type-II and type-III solutions were already obtained by Martins and Guan %
\cite{MG}.

\section{Conclusion}

In this paper we have investigated the regular solutions of the reflection
equations for the vertex models associated to the $D_{n+1}^{(2)}$ affine Lie
algebra. After a systematic study of the functional equations we find that
there are three types of solutions. We call of type-I the $K$-matrix
solutions with $n+2$ free parameters. The type-II are block diagonal
matrices with one free parameter. Finally, the type-III are diagonal
matrices without free parameters which only exist for $n$ even.

The absence of an algebraic method to classify the reflection equation
solutions allows us to believe that our result can be extended in order to
pick up all non-exceptional Lie algebras \cite{Bazhanov, Jimbo}, including
their supersymmetric partners \cite{Schadrikov}.

\vspace{1.0cm}{}

{\bf Acknowledgment:} It is a pleasure to thank  M. J. Martins for
invaluable discussions. This work was supported in part by Funda\c{c}\~{a}o
de Amparo \`{a} Pesquisa do Estado de S\~{a}o Paulo--FAPESP--Brasil and by
Conselho Nacional de Desenvol\-{}vimento--CNPq--Brasil.

\appendix 

\section{The D$_{2}^{(2)}$ Reflection K-matrices}

In this appendix we will present the type-I for the $D_{2}^{(2)}$ model. The
matrix $K_{-}(u)$ has the form%
\begin{equation}
K_{-}=\left( 
\begin{array}{cccc}
k_{11} & k_{12} & k_{13} & k_{14} \\ 
k_{21} & k_{22} & k_{23} & k_{24} \\ 
k_{31} & k_{32} & k_{33} & k_{34} \\ 
k_{41} & k_{42} & k_{43} & k_{44}%
\end{array}%
\right)
\end{equation}%
The elements $k_{11}$, $k_{44}$ and $k_{22}=k_{33}$, $k_{23}$ , $k_{32}$ are
read directly from the $n$ odd solution taking $n=1$ into (\ref{odd.1}),(\ref%
{odd.2}) and (\ref{odd.6}),(\ref{odd.7a}), (\ref{odd.7b}), respectively:%
\begin{eqnarray}
k_{11}(u) &=&\frac{1}{2}\frac{\left[ 2({\rm e}^{2u}-q)(q\beta _{-}^{2}+\beta
_{+}^{2})+(q+1)({\rm e}^{2u}+1)(q\beta _{-}^{2}-\beta _{+}^{2})\right]
k_{14}(u)}{\beta _{14}^{2}\ \sqrt{q}(q+1)({\rm e}^{2u}+1)^{2}}  \nonumber \\
&&+2\frac{k_{14}(u)}{\beta _{14}({\rm e}^{2u}-1)}
\end{eqnarray}

\begin{eqnarray}
\ k_{44}(u)\!\! &=&\frac{1}{2}\!\!\frac{\left[ -2({\rm e}^{2u}-q)(q\beta
_{-}^{2}+\beta _{+}^{2})+(q+1)({\rm e}^{2u}+1)(q\beta _{-}^{2}-\beta
_{+}^{2})\right] {\rm e}^{2u}k_{14}(u)}{\beta _{14}^{2}\ \sqrt{q}(q+1)({\rm e%
}^{2u}+1)^{2}}  \nonumber \\
&&+2\frac{{\rm e}^{2u}k_{14}(u)}{\beta _{14}({\rm e}^{2u}-1)}
\end{eqnarray}%
\begin{equation}
k_{22}(u)=k_{33}(u)=-\frac{1}{2}\left( \frac{q\beta _{-}^{2}-\beta _{+}^{2}}{%
\sqrt{q}(q+1)}\frac{{\rm e}^{2u}+q}{{\rm e}^{2u}+1}+\frac{4\beta _{14}({\rm e%
}^{2u}-q)}{(q-1)({\rm e}^{2u}-1)}\right) \frac{k_{14}(u)}{\beta _{14}^{2}}
\end{equation}%
\begin{equation}
k_{23}(u)=\frac{{\rm e}^{u}}{{\rm e}^{2u}+1}\left( \frac{q\beta
_{-}^{2}+\beta _{+}^{2}}{\sqrt{q}({\rm e}^{2u}+1)}{\rm e}^{u}-\frac{2\sqrt{q}%
\beta _{-}\beta _{+}}{q+1}\right) \frac{k_{14}(u)}{\beta _{14}^{2}}
\end{equation}%
\begin{equation}
k_{32}(u)=\frac{{\rm e}^{u}}{{\rm e}^{2u}+1}\left( \frac{q\beta
_{-}^{2}+\beta _{+}^{2}}{\sqrt{q}({\rm e}^{2u}+1)}{\rm e}^{u}+\frac{2\sqrt{q}%
\beta _{-}\beta _{+}}{q+1}\right) \frac{k_{14}(u)}{\beta _{14}^{2}}
\end{equation}%
where $\beta _{\pm }=\beta _{12}\pm \beta _{13}$.

Due to the indetermination of $\Delta _{l}$ (\ref{str.6}) \ when $n=1$, we
can replace $\Delta _{l}$ into (\ref{str.4}) and (\ref{str.5}) by $\Delta
_{l}^{\prime }$ defined by 
\begin{equation}
\Delta _{l}\rightarrow \Delta _{l}^{\prime }=\left( \frac{q^{2}-1}{%
q^{2}-e^{2u}}\right) \left( \frac{e^{2u}}{1+e^{2u}}\right) \frac{1}{\beta
_{13}(b_{1}^{+}+b_{1}^{-})(b_{4}^{+}+b_{4}^{-})}.
\end{equation}

This replacement allows that the equations (\ref{str.9}--\ref{str.11b}) hold
for the $D_{2}^{(2)}$ model up to a $q$-factor. The result is%
\begin{equation}
k_{41}(u)=\frac{\beta _{21}^{2}}{\beta _{13}^{2}}k_{14}(u),
\end{equation}%
\begin{equation}
k_{12}(u)=\left( \frac{{\rm e}^{u}\beta _{-}+\beta _{+}}{\beta _{14}({\rm e}%
^{2u}+1)}\right) k_{14}(u),\qquad k_{13}(u)=\left( \frac{-{\rm e}^{u}\beta
_{-}+\beta _{+}}{\beta _{14}({\rm e}^{2u}+1)}\right) k_{14}(u),
\end{equation}%
\begin{equation}
k_{21}(u)=\frac{\beta _{21}}{\beta _{13}}k_{13}(u),\qquad k_{31}(u)=\frac{%
\beta _{21}}{\beta _{13}}k_{12}(u),
\end{equation}%
\begin{equation}
k_{42}(u)=\frac{\beta _{21}}{\beta _{13}}k_{34}(u),\qquad k_{43}(u)=\frac{%
\beta _{21}}{\beta _{13}}k_{24}(u).
\end{equation}%
\begin{equation}
k_{24}(u)=\frac{1}{\sqrt{q}}\left( \frac{-q{\rm e}^{-u}\beta _{-}+\beta _{+}%
}{\beta _{14}({\rm e}^{2u}+1)}\right) {\rm e}^{2u}k_{14}(u),\qquad k_{34}(u)=%
\frac{1}{\sqrt{q}}\left( \frac{q{\rm e}^{-u}\beta _{-}+\beta _{+}}{\beta
_{14}({\rm e}^{2u}+1)}\right) {\rm e}^{2u}k_{14}(u),
\end{equation}%
where $\beta _{21}$ is given by 
\begin{equation}
\beta _{21}=\frac{1}{2}\left( \frac{(q-1)(q\beta _{-}^{2}-\beta _{+}^{2})+4%
\sqrt{q}(q+1)\beta _{14}}{(q^{2}-1)\beta _{14}^{2}}\right) \beta _{13}.
\end{equation}

By the choice%
\begin{equation}
k_{14}(u)=\frac{1}{2}\beta _{14}({\rm e}^{2u}-1)
\end{equation}%
we can find $\beta _{11}$ 
\begin{equation}
\beta _{11}=-\frac{2\sqrt{q}}{q+1}\frac{\beta _{12}\beta _{13}}{\beta _{14}}
\end{equation}%
in order to get a regular solution with three free parameters $\beta _{12}$, 
$\beta _{13}$ and $\beta _{14}$.


\begin{thebibliography}{99}
\bibitem{Sklyanin} E. K. Sklyanin, {\em J. Phys A: Math. Gen.} {\bf 21}
(1988) 2375

\bibitem{Alcaraz} F. C. Alcaraz, M. N. Barber, M. T. Batchelor, R. J. Baxter
and G. R. W. Quispel, {\em J. Math. Phys. A: Math. Gen}. {\bf 20} (1987) 6397

\bibitem{Mezincescu1} L. Mezincescu and R. I. Nepomechie, {\em J. Phys. A:
Math. Gen}.{\bf \ 24} (1991) L17

\bibitem{Bazhanov} V. V. Bazhanov, {\em Phys. Lett.} {\bf 159B} (1985) 321; 
{\em Commun. Math. Phys}. {\bf 113} (1987) 471

\bibitem{Schadrikov} V. V. Bazhanov and A. Schadrikov, {\em Theor. Math.
Phys.} {\bf 73} (1987) 402

\bibitem{Cherednik} I. V. Cherednik, {\em Theor. Math. Phys.} {\bf 61}
(1984) 977, {\em Int. J. Mod. Phys}. {\bf A7}, Suppl. {\bf 1B} (1992) 707

\bibitem{deVega} H. J. de Vega and A. Gonz\'{a}lez-Ruiz, {\em J. Phys. A:
Math. Gen.} {\bf 26} (1993) L519; {\em J. Phys. A: Math. Gen}. {\bf 27}
(1994) 6129

\bibitem{Mezincescu2} L. Mezincescu and R. I. Nepomechie, {\em Int. J. Mod.
Phys.} {\bf A6} (1991) 5231

\bibitem{Inami} T. Inami, S. Odake and Y.-Z. Zhang, {\em Nucl. Phys.} {\bf %
B470} (1996) 419

\bibitem{Batchelor} M. T. Batchelor, V. Fridkin, A. Kuniba and Y. K. Zhou, 
{\em Phys. Lett.} {\bf B376} (1996) 266

\bibitem{Lima} A. Lima-Santos, {\em Nucl. Phys.} {\bf B558} (1999) 637

\bibitem{Liu} Cong-xin Liu, Guo-xing Ju, Shi-kun Wang and Ke Wu, {\em J.
Phys. }{\bf A32} (1999) 3505

\bibitem{MG} M. J. Martins and X.-W. Guan, {\em Nucl. Phys.} {\bf B583}
(2000) 721

\bibitem{Jimbo} M. Jimbo, {\em Commun. Math. Phys}. {\bf 102} (1986) 537

\bibitem{Cuerno} R. Cuerno and A. Gonzales-Ruiz, {\em J. Phys. A: Math. Gen.}
{\bf 26} (1993) L605
\end{thebibliography}
\end{document}